\documentclass[aps,prl,reprint,superscriptaddress,showpacs]{revtex4-1}

\usepackage{color}

\usepackage{graphicx}
\usepackage{booktabs}
\usepackage{bm}
\usepackage{longtable}
\usepackage{amsmath}
\usepackage{dcolumn}
\usepackage{placeins}

\usepackage{siunitx}

\begin{document}

\title{Imaging of material defects with a radio-frequency atomic magnetometer}

\author{P. Bevington} 
\affiliation{National Physical Laboratory, Hampton Road, Teddington, TW11 0LW, United Kingdom}
\affiliation{Department of Physics, University of Strathclyde, Glasgow G4 0NG, United Kingdom}
\author{R. Gartman} 
\affiliation{National Physical Laboratory, Hampton Road, Teddington, TW11 0LW, United Kingdom}
\author{W. Chalupczak}
\affiliation{National Physical Laboratory, Hampton Road, Teddington, TW11 0LW, United Kingdom}

\date{\today}

\begin{abstract}
Non-destructive eddy current testing of defects in metal plates using the magnetic resonance signal of a radio-frequency atomic magnetometer is demonstrated. The shape and amplitude of the spatial profile of signal features that correspond to defects are explored. By comparing numerical and experimental results on a series of benchmark aluminium plates we demonstrate a robust process for determining defect dimensions. In particular, we show that the observed images represent the spatial distribution of the secondary field created by eddy currents in the sample. We also demonstrate that the amplitude and phase contrast of the observed profiles enables us to reliably measure defect depth.
\end{abstract}


\maketitle

Accurate identification of structural flaws, such as corrosion in pipes or cracks in an aerofoil, has obvious benefits across many industries. In particular, the global cost of corrosion is estimated to be USD 2.5 trillion every year \cite{IMPACT2016}. Over 20$\%$ of the major oil and gas accidents reported within the EU since 1984 have been associated with corrosion under insulation \cite{OGTC2018}. Monitoring the sample response to an oscillating magnetic primary field ($\vec{B}$), provides a method of non-destructive testing (NDT) and detection of anomalies. Sample response, i.e. an oscillating secondary magnetic field ($\vec{b}$), is created by eddy currents in samples with high electrical conductivity, and magnetisation in ones with high magnetic permeability \cite{Footnote1}. The response  carries information about the conductivity, permittivity and permeability of the object, and these material properties can be mapped via the measurement of $\vec{b}$ \cite{Griffiths2001}.

Most eddy current NDT systems use an rf coil to generate $\vec{b}$, but the material response can be detected in two ways, namely by measurement of coil inductance or with a magnetic sensor. Measurements based on coil inductance monitoring \cite{Griffiths2001, Auld1999, Perez2004, Sophian2017} benefit from simple instrumentation, but suffer from a decrease in signal sensitivity at low frequencies. Magnetic sensors offer higher sensitivity at low-frequencies and better spatial resolution \cite{Ripka2010}. This category of sensors includes giant magnetoresistance (GMR) magnetometers \cite{Dogaru2000, Dogaru2001, Ripka2010}, superconducting quantum interference devices (SQUIDs) \cite{Krause2002, Storm2017} and atomic magnetometers \cite{Wickenbrock2014, Deans2016, Wickenbrock2016, Deans2017}.
Atomic magnetometers are an attractive sensor type as they have sensitivities approaching those of GMR and SQUID sensors, can operate in an unshielded environment \cite{Savukov2007, Belfi2010, Bevilacqua2016}, do not require cryogenics, and have few restrictions of miniaturisation.

\begin{figure}[htbp]
\includegraphics[width=\columnwidth]{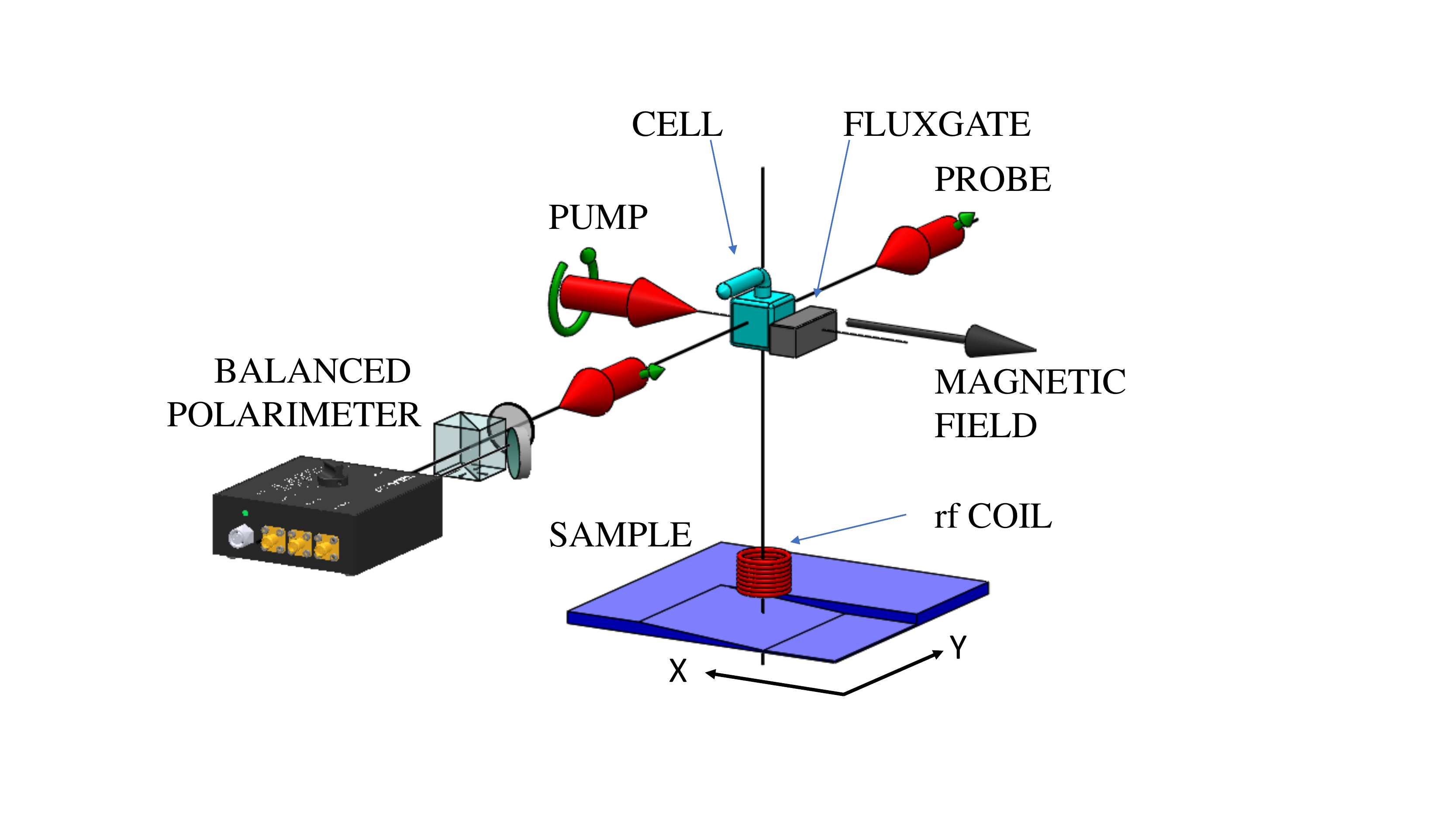}
\caption{Main components of the experimental setup. The oscillating primary magnetic field is generated by rf coil. The secondary magnetic field is produced by eddy currents excited in a sample (Al plate with recess) by the primary field. The atomic magnetometer signal monitors contributions from the primary as well as the secondary magnetic field. }\label{fig:Setup}
\end{figure} 
 
The ability to extract information about defects such as cracks, bends and pipe thinning from measurement signals lies at the heart of NDT. Therefore, understanding how the observed signal is created and linked to the anomaly is one of the first issues that needs to be addressed when developing an NDT system. In this paper we explore how the spatial variation in the signals recorded by an rf magnetometer correspond to the physical dimensions of structural defects in aluminium (Al) plates. Non-magnetic samples have been chosen to minimise changes in the local field detected by the atomic magnetometer. However, while we examine the spatial profiles generated by the thinning of Al targets, similar profiles have been observed in the case of ferromagnetic (carbon steel) plates \cite{Bevington2018}. The arguments presented here could, with some modifications, be extrapolated to measurements involving high magnetic permeability materials. In addition to the characterisation of the measurement signal's spatial profile, we calibrate measurements of sample thinning by monitoring the change of the rf signal phase/amplitude.
This work presents a specific geometry, where the atomic sensor and coil are on the same side of the sample and are operated in an unshielded environment, representing a real-world sensing environment.

The measurement signal comes from the phase and amplitude change in the rf resonance spectra registered by an rf atomic magnetometer as a conductive sample is moved under the rf coil  (Fig. \ref{fig:Setup}).
Since the experimental setup is similar to that described in \cite{Chalupczak2012, Bevington2018} only essential components will be discussed here. Room temperature caesium vapour (atomic density $n_{\text{Cs}}=\SI{3.3e10}{\per\centi\meter\cubed})$ atoms are optically pumped in to the stretched state ($F= 4, m_{F} = 4$) with a circularly polarised laser locked to the Cs $6\,^2$S$_{1/2}$ F=3$\rightarrow{}6\,^2$P$_{3/2}$ F'=2 transition (D2 line, $\SI{852}{\nano\meter}$) propagating along the bias magnetic field. 
The probe beam ($\SI{30}{\micro\watt}$) is $\SI{580}{\mega\hertz}$ blue shifted from the $6\,^2$S$_{1/2}$ F=4$\rightarrow{}6\,^2$P$_{3/2}$ F'=5 transition via phase-offset-locking to the pump beam. Coherent spin precession of the Cs atoms is coupled to the polarisation of the probe beam (Faraday rotation) and is detected with a balanced polarimeter. This signal is then demodulated at radio-frequency by a lock-in amplifier (SRS 865).
This work is carried out in a magnetically unshielded environment, where static fields along the $y$ and $z$ directions are nulled and a bias field is created by three pairs of nested, orthogonal, square Helmholtz coils, with dimensions $\SI{1}{\meter}$, $\SI{0.94}{\meter}$ and $\SI{0.88}{\meter}$ respectively. The operating frequency of the system (i.e. the magnetic resonance frequency) is set to $\SI{12.6}{\kilo\hertz}$ by the $x$ component of the bias magnetic field (Fig. \ref{fig:Setup}).
The rf coil is a 1000 turn coil with $\SI{0.02}{\milli\meter}$ wire, wound on a $\SI{2}{\milli\meter}$ plastic core (inside diameter) and with a $\SI{4}{\milli\meter}$ width (outside diameter) and a $\SI{10}{\milli\meter}$ length. It is driven by an internally referenced rf output from the lock-in amplifier. The samples are fixed to a 2D, computer controlled translation stage that has a minimum step of $\SI{0.184}{\milli\meter}$. The sample is located approximately $\SI{30}{\centi\meter}$ from the cell and the coil is placed $\SI{1}{\milli\meter}$-$\SI{2}{\milli\meter}$ above the sample, on the same axis as the cell.

\begin{figure}[htbp]
\includegraphics[width=\columnwidth]{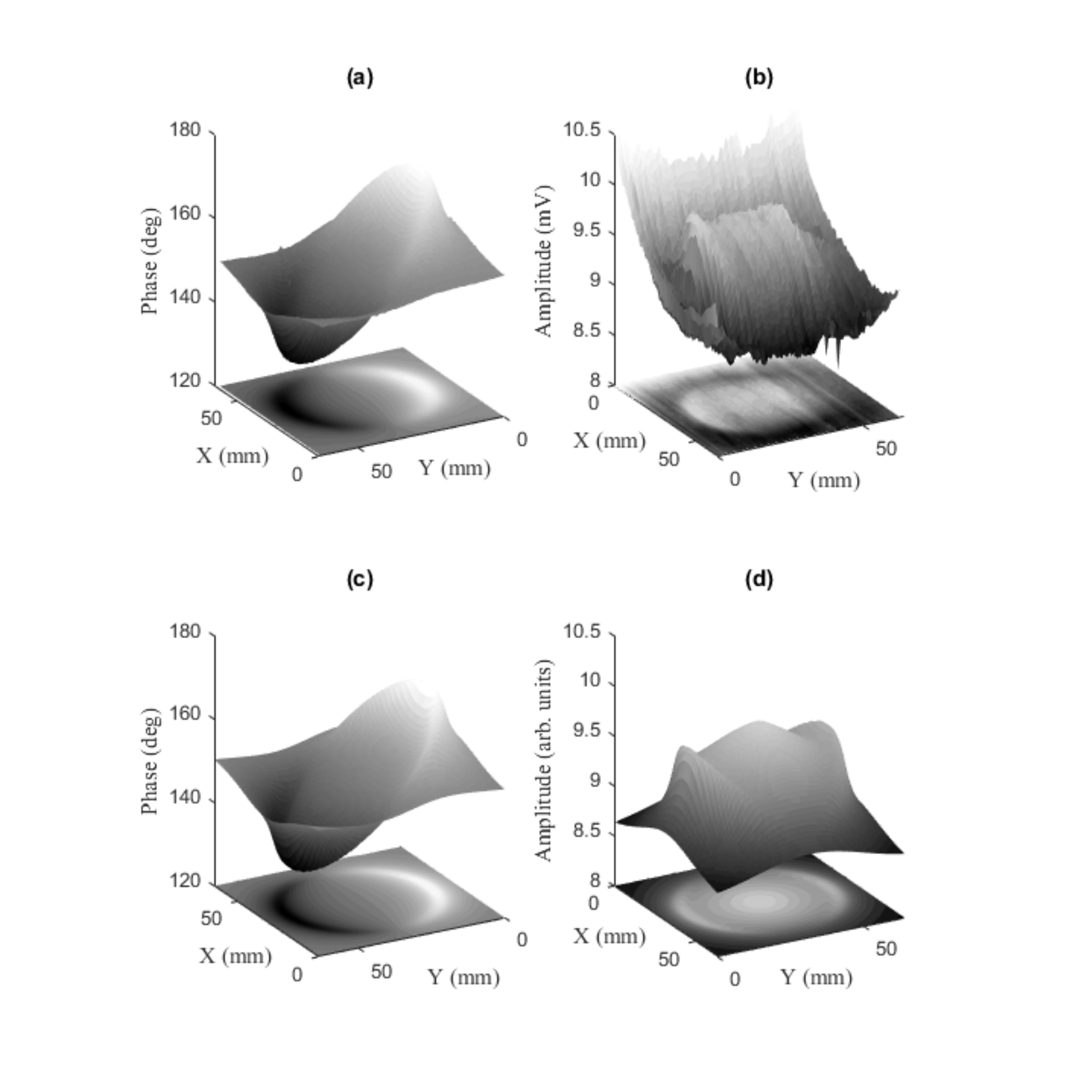}
\caption{The measured change in signal phase (a) and amplitude (b) of the rf resonance over a $64\times\SI{64}{\milli\meter\squared}$ area of a \SI{6}{\milli\meter} thick Al plate containing a \SI{48}{\milli\meter} diameter recess that is \SI{2.4}{\milli\meter} deep. Below, results of simulation of the phase (c) and amplitude (d).} \label{fig:Edge}
\end{figure}

Figure \ref{fig:Edge} (a)/(b) shows the results of the scanned $64\times\SI{64}{\milli\meter\squared}$ area of an Al plate with a defect in the form of a recess ($\SI{48}{\milli\meter}$ diameter, \SI{2.4}{\milli\meter} deep). Each pixel of the image represents the phase [Fig.~\ref{fig:Edge} (a)] or amplitude [Fig.~\ref{fig:Edge} (b)] of the rf resonance profile recorded by scanning the rf frequency through the magnetic resonance. These parameters are extracted through fitting of a Lorentzian and dispersive profile to the rf resonance line shapes. Although both plots contain traces of the sample thinning, the spatial signatures of the recess have different characteristics. The feature presented in Fig.~\ref{fig:Edge} (a) shows that the phase changes as the rf coil scans over the falling and rising edges. The phase change measurement is most sensitive to the edges parallel to the bias field. Signatures from the parallel edges are also present in the amplitude data, Fig.~\ref{fig:Edge} (b), but are superimposed on top of an increase in signal recorded over the whole recess.

\begin{figure}[htbp]
\includegraphics[width=\columnwidth]{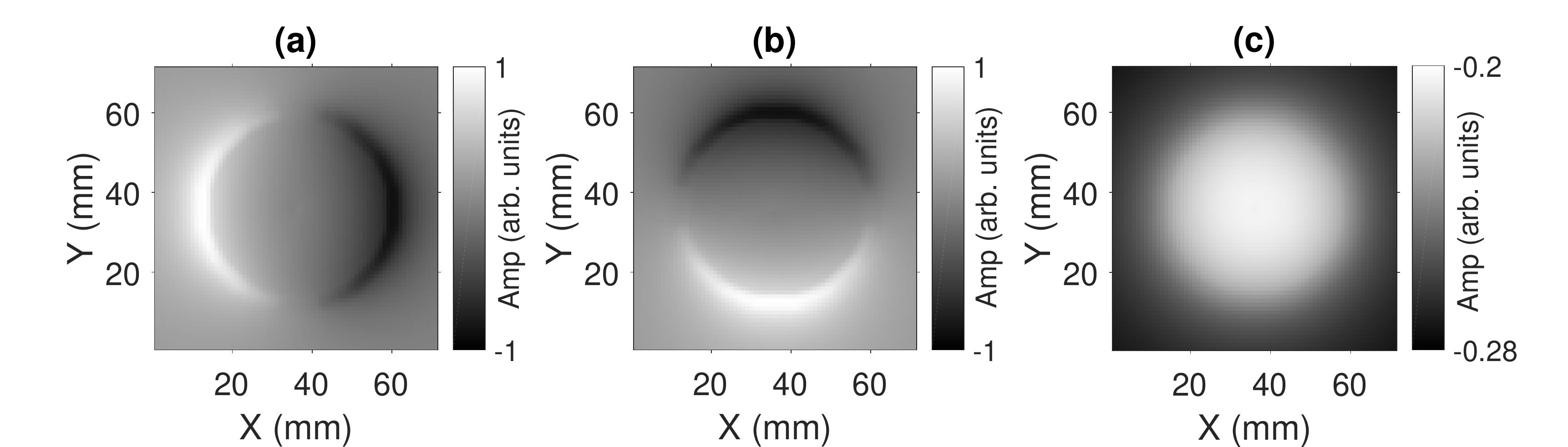}
\caption{Calculated $\vec{b}$ field components generated in a $64\times\SI{64}{\milli\meter\squared}$ conductive area, a with \SI{48}{\milli\meter} diameter recesses, (a), (b), and (c) refer to $x$, $y$ and $z$ component respectively. }\label{fig:Modell}
\end{figure}

In order to gain insight into the recorded images, we created a simple 2D model based on Faraday's law, by calculating the coupling between the primary field $\vec{B}$ and the conductive sample containing an inhomogeneity. We model the spatial distribution of $\vec{B}$ with a step function that describes the rf coil diameter. The secondary field, $\vec{b}$, changes linearly within the boundaries of the step function and decreases inversely with the distance outside of it. The eddy currents form closed loops that follow the path of least resistance.
In the case of a uniform conductor surface, $\vec{b}$ will be produced parallel to the surface normal. However, the presence of inhomogeneities breaks the eddy currents symmetry and can change the orientation of $\vec{b}$. Figure \ref{fig:Modell} shows the components of $\vec{b}$ generated in a conductive sample in the presence of a \SI{48}{\milli\meter} diameter recess. Since the direction of $\vec{b}$ depends on the relative position of the induced eddy currents and the recess boundaries, $\vec{b}$ has opposite signs for rising and falling edges [Fig. \ref{fig:Modell} (a, b)]. The components of $\vec{b}$ parallel to the conductor surface show complementary signatures of the recess (i.e. Fig. \ref{fig:Modell} (a)/(b) shows the change of the field sign due to a presence of part of the recess edges parallel to the $y/ x$ direction).
The component of $\vec{b}$ orthogonal to the conductor surface decreases over the recess area, Fig. \ref{fig:Modell} (c).

While Fig. \ref{fig:Modell} only shows the components of the secondary field, the signal measured by the magnetometer is a mixture of both the primary, $\vec{B}$, and secondary field, $\vec{b}$. The lock-in signal amplitude R=$\sqrt{( (B_z + b_z)^2+(B_y + b_y)^2)}$ and phase $\phi = \arctan(\frac{B_y + b_y}{B_z + b_z})$ correspond to the strength and orientation of the rf field component projected on the $yz$ plain; where $B_z + b_z$ and $ B_y + b_y$ are the two quadrature components of the rf signal. In our experiment $B_y \simeq 0$ and $B_z \simeq const \gg b_z$, hence $B_z + b_z \gg B_y + b_y$. Therefore we can make the approximation that $R \simeq |B_z + b_z|$ and $\phi  \simeq b_y/(B_z+b_y)$ which leads to mapping of  $b_z$ onto $R \propto b_z$ and $b_y$ onto $\phi \propto b_y$. The latter approximation comes from the observation that for $\lvert \vec{B} \rvert \gg \lvert \vec{b} \rvert$, the function $\phi$ depends more strongly on $b_y$ ( $d\phi/db_y \propto B_z ^{-1}$) than $b_z$ ($ d\phi/db_z\propto B_z ^{-2}$). It can be concluded that the presence of a strong $\vec{B}$ allows mapping of one horizontal component and the vertical component of $\vec{b}$ onto the amplitude and phase of the rf signal, as is shown in Figs. \ref{fig:Edge} and \ref{fig:Modell}.
Figure \ref{fig:Edge} (c)/(d) shows the results of the modelled signal with contributions from both the primary and secondary fields. Full reconstruction of $\vec{b}$ would require rotation of the sensor axis to measure the remaining horizontal component.

\begin{figure}[htbp]
\includegraphics[width=\columnwidth]{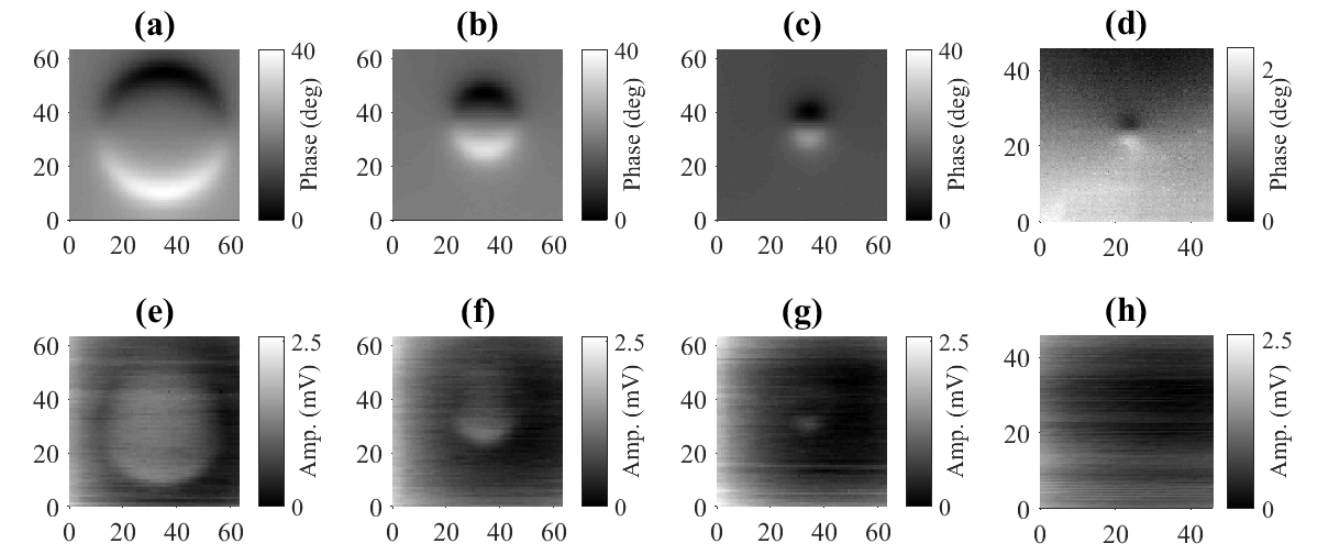}
\caption{The change in phase (a, b, c, d) and amplitude (e, f, g, h) of rf resonance over a $64\times\SI{64}{\milli\meter\squared}$ area of \SI{6}{\milli\meter} thick Al plates with \SI{48}{\milli\meter} (a, e), \SI{24}{\milli\meter} (b, f), \SI{12}{\milli\meter} (c, g) and \SI{2}{\milli\meter} (d, h) diameter recesses that are \SI{2.4}{\milli\meter} deep.} \label{fig:RecessDiameter}
\end{figure}

\begin{figure}[htbp]
\includegraphics[width=\columnwidth]{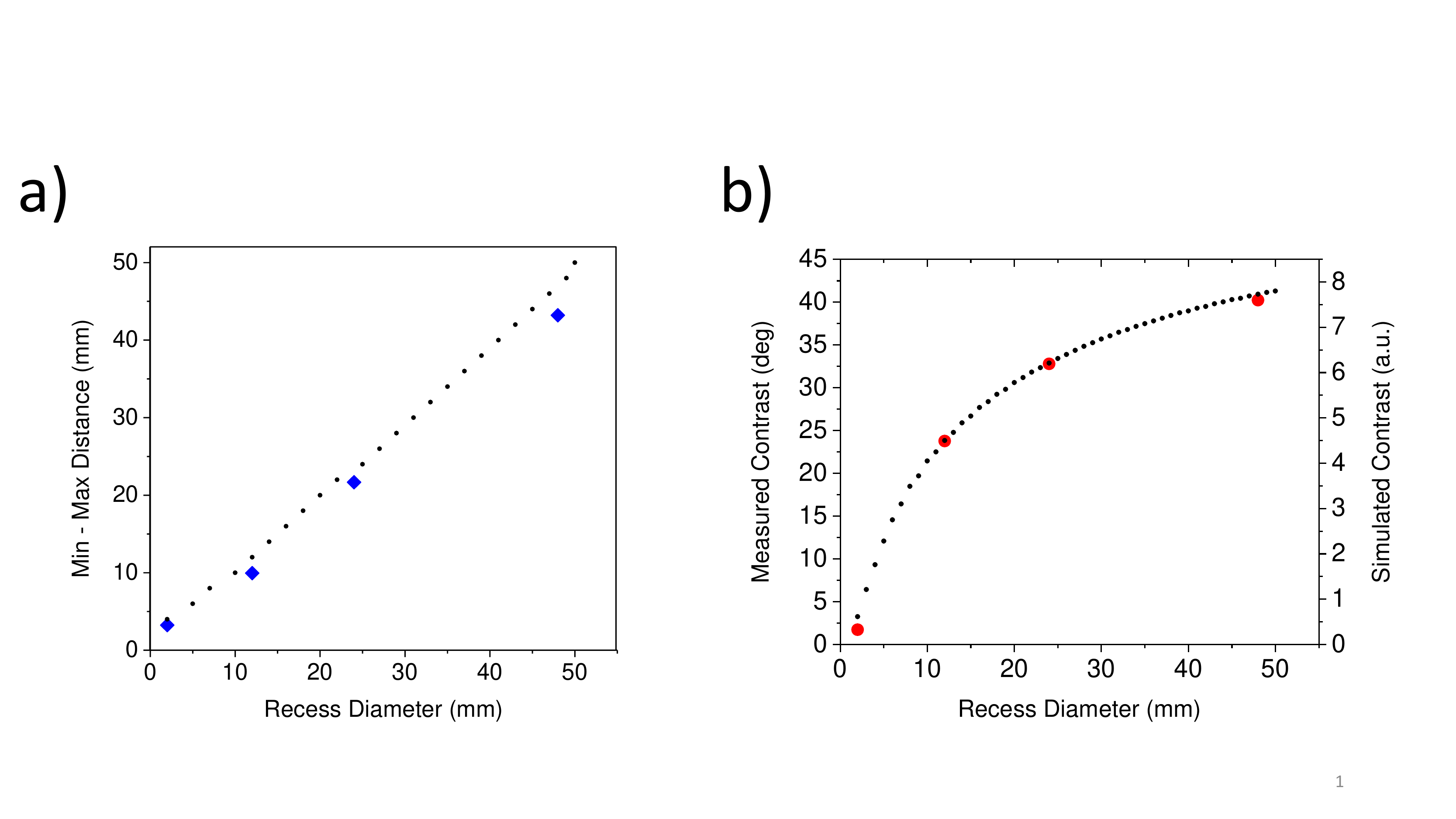}
\caption{Dependence of measured feature size (blue diamonds) and phase contrast (red points) on actual size of circular recesses. Dashed lines represent the modelling.}\label{fig:RecessesDiameterContrast}
\end{figure}

The minimum spatial extent of the profile representing the defect (recess) is defined by the coil diameter, lift off distance, operating rf frequency, and conductivity of the sample. The radial dimension of the rf coil defines the $\vec{B}$ spatial distribution and the size of the region containing the greatest density of eddy currents \cite{Dogaru2001}. There are two regimes in eddy current testing that are defined by the ratio between the dimensions of the rf coil and the size of the defect. When the defect is significantly larger than the coil, the observed profile in the image represents the spatial extent of the defect. If the defect is smaller than the coil size, the image represents the map of the field generated by the rf coil \cite{Auld1999}. We explore these two regimes by recording the phase images of circular recesses of decreasing diameter in aluminium Al plates (Fig.~\ref{fig:RecessDiameter}). While the diameters of the features shown in Fig. \ref{fig:RecessDiameter} (a-c) follow the diameters of the actual recesses, the diameter of the profile shown in (d), which represent the phase image of a \SI{2}{\milli\meter} recess, is defined by the \SI{4}{\milli\meter} diameter of the rf coil, as shown in Fig. \ref{fig:RecessesDiameterContrast} (a). 
We have observed the same behaviour with even smaller recesses. This confirms the initial assumption that the spatial resolution of our measurement is limited by the size of the coil and is not restricted by the sensor (vapour cell) dimensions. It is worth pointing out that while the \SI{2}{\milli\meter} recess is clearly visible in the phase image it can't be identified in the amplitude image.
In the case of recesses with small diameters, the features in the phase image that represent edges overlap and the actual value of the phase contrast becomes a function of the recess diameter, as shown in Fig. \ref{fig:RecessesDiameterContrast} (b).

\begin{figure}[htbp]
\includegraphics[width=\columnwidth]{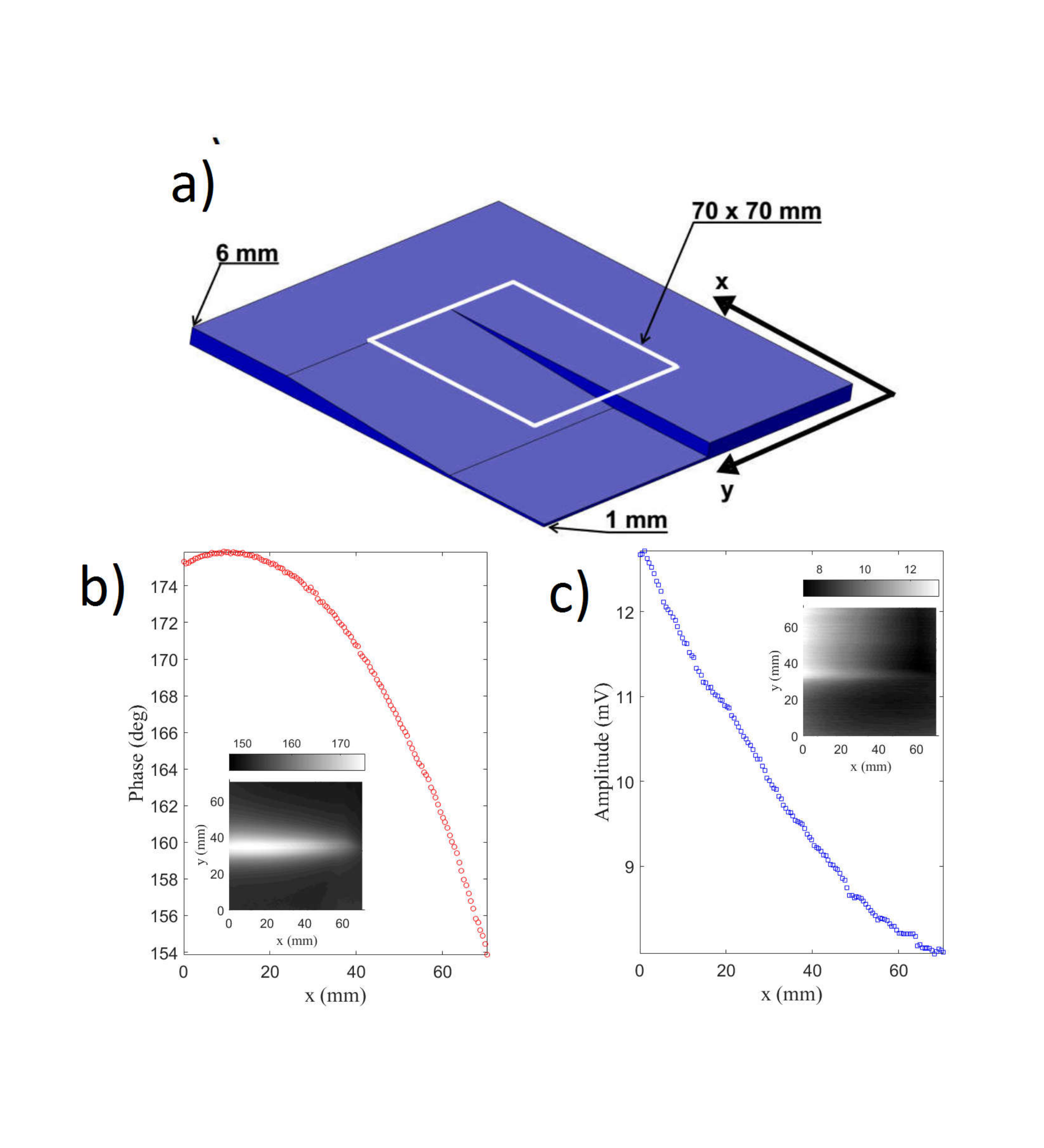}
\caption{(a) Drawing of the aluminium plate used for a contrast calibration. White contour indicates the area of the scan. The change in phase (b) and amplitude (c) of rf resonance over a recess edge (main plot)/ a $\SI{70}\times\SI{70}{\milli\meter\squared}$ area of a \SI{6}{\milli\meter} Al plates (inset).}\label{fig:Recess}
\end{figure}

To verify the dependence of the phase and amplitude contrast on the depth of the recess we have recorded phase and amplitude images generated by a \SI{6}{\milli\meter} Al plate with a single recess, whose depth changes from \SI{0}{\milli\meter} to \SI{5}{\milli\meter} (scanned area marked with white square in Figure \ref{fig:Recess} (a)). Scanning the plate along the x-axis is equivalent to monitoring the signal response to a continuous change in the depth of the recess. The presence of only one recess in the scanned area ensures that the observed amplitude and phase changes are not affected by other inhomogeneities in the sample. Figure \ref{fig:Recess} shows the phase (b) and amplitude (c) change of the rf signal recorded over the recess. It indicates a linear change of the signal amplitude and a sinusoidal change of the signal phase with the depth of the recess. There is a change in a slope of linear dependence in amplitude variation for x $\le \SI{10}{\milli\meter}$ and x $\ge \SI{60}{\milli\meter}$ in Fig. \ref{fig:Recess} c. The central part of the plot comes from scan over the region with monotonic change of the recess height. The regions  mentioned above include contributions from the flat parts outside scanning range due to finite size of the rf coil.

To conclude, monitoring the amplitude and phase changes of the rf resonance over the conductive sample provides a sensitive tool for the detection of a material defect. Implementation of atomic magnetometers in eddy current imaging is particularly interesting for studies of ferromagnetic samples at low operating frequencies, enabling penetration through the sample surface or under insulation barriers (e.g. corrosion under insulation). We have demonstrated that the phase and amplitude images could be used for estimations of sample thinning, although specific thickness estimation needs to take into account the size of the inhomogeneity. The tunability of the rf atomic magnetometer not only allows for the change of the penetration depth but also the choice of an operating frequency range free from external interferences.

\begin{acknowledgements}
This work was funded by the Innovate UK Energy Game Changer programme (IUK 132437). 
\end{acknowledgements}


\begin{thebibliography}{10}
\bibitem{Footnote1} In ferromagnetic samples, the oscillation in magnetisation is much larger than the magnetic field generated by eddy currents.  Measurement of the driven eddy current response is therefore aided by saturation of the magnetisation.
\end{thebibliography}
\end{document}